\documentclass[aps,prb,twocolumn]{revtex4}
\usepackage{amssymb}
\usepackage{amsfonts}
\usepackage{amsmath}
\usepackage{graphicx}

\begin{document}

\title{Fishtail effect and the vortex phase diagram of single
crystal Ba$_{0.6}$K$_{0.4}$Fe$_{2}$As$_{2}$}

\author{Huan Yang, Huiqian Luo, Zhaosheng Wang and Hai-Hu Wen$^*$}

\address{National
Laboratory for Superconductivity, Institute of Physics and Beijing
National Laboratory for Condensed Matter Physics, Chinese Academy of
Sciences, P.O.~Box 603, Beijing 100190, P.~R.~China}

\begin{abstract}
By measuring the magnetization hysteresis loops of superconducting
Ba$_{0.6}$K$_{0.4}$Fe$_{2}$As$_{2}$ single crystals, we obtained the
high upper critical field and large current carrying ability which
point to optimistic applications. The fishtail (or second peak)
effect is also found in the material, and the position of the vortex
pinning force shows a maximum at 1/3 of the reduced field, being
consistent with the picture of vortex pinning by small size normal
cores in the sample. Together with the resistive measurements, for
the first time the vortex phase diagram is obtained for
superconductor Ba$_{0.6}$K$_{0.4}$Fe$_{2}$As$_{2}$.
\end{abstract}

\maketitle

\newpage

The discovery of high temperature superconductivity in
LaFeAsO$_{1-x}$F$_x$ has stimulated enormous interests in the field
of superconductivity\cite{LaOFFeAs}. The family of
REFeAsO$_{1-x}$F$_x$ (FeAs-1111 phase)exhibits quite high
superconducting transition temperatures with the maximum
$T_\mathrm{c}=55\;$K for RE=Sm \cite{SmF_RZA} in electron doped
region, as well as $25\;$K in hole doped La$_{1-x}$Sr$_x$FeAsO
\cite{LaSr_Wen}. Recently, BaFe$_2$As$_2$ compound with new
structure (FeAs-122 structure) was reported\cite{BaFeAs}, and by
replacing the alkaline earth elements (Ba, Ca, and Sr) with alkali
elements, superconductivity with maximum $T_\mathrm{c}=38\;$K was
discovered. \cite{BK1,BK2,BK3} The iron based superconductors have
quite high upper critical fields\cite{LaHc2,NdSC1} which indicates
encouraging applications. Compared with the cuprate system, they
have both differences and similarities. Concerning the differences
between Fe-based and Cu-based superconductors, they involve mainly
the symmetry of the order parameter: the former may have an $s$-wave
symmetry, while the latter is $d$-wave like. Another important
difference is that the iron based superconductors have smaller
anisotropy\cite{WenReview}, for example, about 5-6 near
$T_\mathrm{c}$ in NdFeAsO$_{0.82}$F$_{0.18}$\cite{NdJia}. The
anisotropy, as in the case of YBa$_2$Cu$_3$O$_{7-\delta}$ (YBCO), is
about 7-20 near the optimal doping and strongly dependent on the
doping level of the sample\cite{WenReview}. The research on the
vortex dynamics is a useful tool to make the judgment, and recent
works suggest that vortex properties in the two materials seem to be
similar to each other. \cite{NdJia,SmYang,NdProzorov,NdDou}

For some superconductors, the critical current density obtained from
the magnetization hysteresis loops (MHL) increases with the magnetic
field after the first peak of penetration field. This is the
so-called fishtail effect or second peak effect. This feature has
been observed in the clean and high quality single crystals of
cuprate superconductors; while in the low $T_\mathrm{c}$
superconductors, e.g. Nb$_3$Sn\cite{peakNb3Sn},
MgB$_2$\cite{peakMgB2}, etc, peak effect has been observed near the
upper critical field $H_\mathrm{c2}$ which may be induced by a
different mechanism. The anomalous second peak appears at different
fields for REBa$_2$Cu$_3$O$_{7-\delta}$ (REBCO) bulks (RE=rare earth
element) at different temperatures\cite{RBaCuO}, however, the peak
position is temperature independent for Bi-based and Tl-based
cuprate\cite{BiTl}. They may have different origins in REBCO and in
Bi- or Tl-based cuprates, and have not got final consensus. The
second peak effect has been observed and reported in polycrystalline
samples of the FeAs-1111 phase\cite{LaPeak,NdPeak,SmPeak1,SmPeak2}.
However, it has not been reported in the FeAs-122 phase either in
polycrystalline or single crystal samples. The effect thus deserves
a detailed investigation on single crystals.

The single crystal Ba$_{0.6}$K$_{0.4}$Fe$_{2}$As$_{2}$ samples were
grown by the FeAs self-flux method, and the detailed preparation
process was given elsewhere \cite{SampleLuoHQ}. The sample in this
work was shaped into a rectangle with the dimensions of
$1.054\;\mathrm{mm\,(length)}\times0.548\;\mathrm{mm\,
(width)}\times0.038\;\mathrm{mm\,(thickness)}$ for both the magnetic
and resistive measurements. The measurements were carried out on a
physical property measurement system (PPMS, Quantum Design) with the
magnetic field up to 9$\;$T ($H\parallel c$). The temperature
stabilization is better than 0.1\% at fixed temperatures. The
magnetic properties were measured by the sensitive vibrating sample
magnetometer (VSM) at the vibrating frequency of $40\;$Hz with the
resolution better than $1\times10^{-6}\;$emu. The advantage of this
technique is that the data acquisition is very fast with a quite
good resolution for magnetization. The magnetic field sweeping rate
for the MHL was $100\;$Oe/s.

\begin{figure}[h]
\includegraphics[width=7cm]{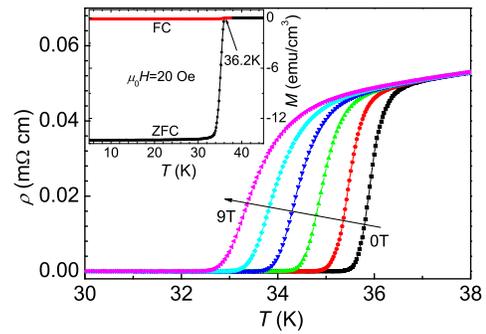}
\caption{(Color online) Temperature dependence of resistivity at
various fields of 0, 1, 3, 5, 7, 9$\;$T (from right to left) of the
Ba$_{0.6}$K$_{0.4}$Fe$_{2}$As$_{2}$ single crystal. The inset shows
temperature dependence of the superconducting diamagnetic moment
measured in the ZFC and FC processes at a field of 20$\;$Oe.}
\label{fig1}
\end{figure}

Fig.~1 shows the temperature dependence of resistivity at various
magnetic fields. By taking a criterion of $95\%\rho_\mathrm{n}$, the
onset transition temperature at zero temperature is about $36.5\;$K,
while the zero-resistance temperature is about $35.2\;$K. The
diamagnetic transition of the sample is shown in the inset of
Fig.~1, which was measured in the field-cooled (FC) and
zero-field-cooled (ZFC) processes at a DC magnetic field of
$20\;$Oe. The ZFC curve shows perfect diamagnetism in the low
temperature region and the sharp transition with the onset
temperature $T_\mathrm{c}=36.2\;$K and transition width $1.6\;$K.
Both the magnetic and the resistive measurement indicate good
quality of the sample which warrants a further detailed
investigation on vortex dynamics. Using the criterion of upper
critical field is taken from the resistive transition by
$95\%\rho_\mathrm{n}$, a huge slop
$\mathrm{d}H_\mathrm{c2}/\mathrm{d}T=-6.89\;$T/K is obtained. The
rough estimation of $H_\mathrm{c2}$ at zero temperature is $174\;$T
by using the Werthamer-Helfand-Hohenberg (WHH) formula\cite{WHH}
$H_\mathrm{c2}(0)=-0.69\mathrm{d}H_\mathrm{c2}(T)/\mathrm{d}T|_{T_\mathrm{c}}T_\mathrm{c}$.

\begin{figure}[h]
\includegraphics[width=7cm]{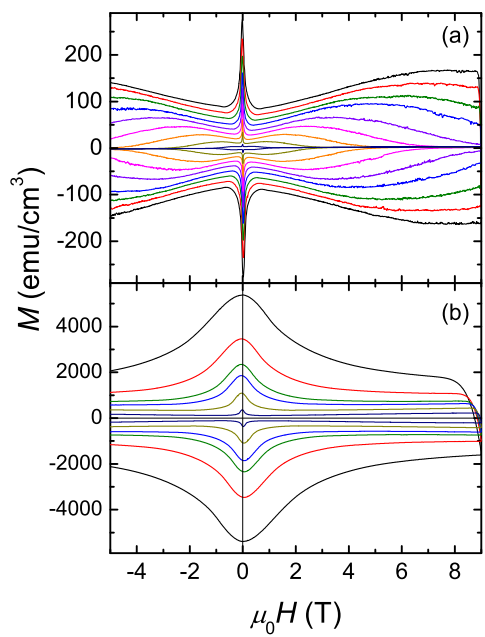}
\caption{(Color online) Magnetization hysteresis loops at the
temperatures  (a) from 27 to 35$\;$K with an interval of 1$\;$K, (b)
2, 5, 10, 15, 20, and 25$\;$K,. Fishtail effect appears in the high
field region.} \label{fig2}
\end{figure}

In Fig.~2, the MHL measured at different temperatures from 2$\;$K to
35$\;$K are presented. The symmetric curves suggest that the bulk
pinning instead of the surface barrier dominates in the sample. The
second peak can be easily observed in Fig.~2~(a) at $T>27\;$K. With
the decreasing of $T$, the peak moves to the high field and finally
goes beyond the field region with the maximum value $9\;$T as shown
in Fig.~2~(b). In the region between the valley and the peak, the
width of the irreversible magnetization $\Delta M$ increases with
increasing of the magnetic field, which shows a clear fishtail
effect. A monotonous dropping down of the peak position with
increasing temperature can be found in Fig.~2~(b), which suggests
that the second peak effect may have the same mechanism as that in
REBCO\cite{RBaCuO}. Generally, the field dependence of critical
current density $J_\mathrm{c}$ can be calculated from the MHL based
on the Bean critical state model\cite{Bean} $J_\mathrm{c}=20\Delta
M/[w(1-w/3l)]$, where $\Delta M$ was measured in emu/cm$^3$, and the
length $l$, the width $w$ of the sample ($w<l$) were measured in cm.
Fig.~3 shows the field dependence of $J_\mathrm{c}$. The calculated
$J_\mathrm{c}$ at 2$\;$K and $0\;$T reaches
4.7$\times10^6\;$A/cm$^2$, and $J_\mathrm{c}$ at 5$\;$K remains more
than 1$\times10^6\;$A/cm$^2$ at $9\;$T. So the material shows the
good current carrying ability.

\begin{figure}[h]
\includegraphics[width=7cm]{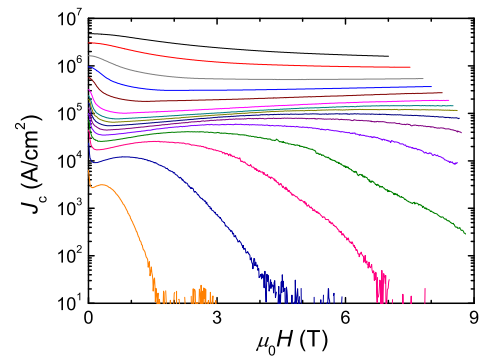}
\caption{(Color online) Field dependence of the calculated critical
current density from the Bean critical state model at the
temperatures corresponding to all the curves in Fig.~2.}
\label{fig3}
\end{figure}

In order to find the pinning mechanism of the vortices associating
with the fishtail effect, the pinning force density
$F_\mathrm{p}\propto J_\mathrm{c}H$ can provide much information. As
proposed by Dew-Hughes\cite{FpvsH}, the pinning force density
$F_\mathrm{p}$ should be proportional to $h(1-h)^2$ as the pinning
is induced by the small size normal cores. Here $h$ is defined as
$h=H/H_\mathrm{irr}$ where $H_\mathrm{irr}$ is obtained from the
zero value of $J_\mathrm{c}$ in $J_\mathrm{c}$-$\mu_0H$ curves. The
expression above yields a maximum value at $h\approx1/3$. Fig.~4
presents the relationship between normalized pinning force density
$F_\mathrm{p}/F_\mathrm{p}^\mathrm{max}$ and $h=H/H_\mathrm{irr}$ of
the three curves at high temperatures near $T_\mathrm{c}$. Three
curves overlap well with the peak position locating at around $h\sim
0.33$. So the fishtail effect in high temperature region of this
sample is from the small size normal cores, which is similar as the
results in YBCO.\cite{WenAPL}

\begin{figure}[h]
\includegraphics[width=7cm]{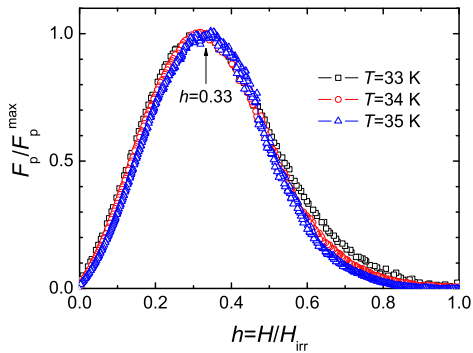}
\caption{(Color online) The correlation between the normalized
pinning force $F_\mathrm{p}/F_\mathrm{p}^\mathrm{max}$ and the
reduced field $h=H/H_\mathrm{irr}$. The maximum peak locates at the
position $h\sim0.33$ which is predicted by the theory for the small
size normal core pinning.} \label{fig4}
\end{figure}

In Fig.~5, we present the vortex phase diagram of the material. For
the magnetic measurement, three characteristic fields are confirmed
as shown by the solid symbols in the inset of Fig.~5.
$H_\mathrm{min}$ and $H_\mathrm{sp}$ locate at the valley and the
peak of the curve, respectively; the irreversibility field
$H_\mathrm{irr}$ is determined by taking a criterion of
$10\;\mathrm{A/cm}^2$. The irreversibility field is also obtained
from the zero-resistance temperature in the $\rho$-$T$ curves shown
as open squares in Fig.~5; and the upper critical field
$H_\mathrm{c2}$ with $95\%\rho_\mathrm{n}$ are shown as the open
circles. The irreversibility lines determined by magnetic and
resistive measurements are slightly different. As
$H_\mathrm{min}$-$T$ and $H_\mathrm{sp}$-$T$ curves have more
information rather than the linear behavior, we try to find the
functional relationship. The two curves are well fitted by the
expressions
$H_\mathrm{min}(T)=H_\mathrm{min}(0)(1-T/T_\mathrm{c})^{1.3}$ with
$H_\mathrm{min}(0)=4.0\;$T and
$H_\mathrm{sp}(T)=H_\mathrm{sp}(0)(1-T/T_\mathrm{c})^{1.5}$ with
$H_\mathrm{sp}(0)=62.9\;$T. The same power-law behavior between
$H_\mathrm{sp}$ and $T$ is found in YBCO \cite{YBCOPeak} and the
second peak effect is strongly influenced by the oxygen deficiency
there. Here in iron based superconductors, the local magnetic
moments may form the small size normal cores, and it may be a
possible reason of the second peak effect. However, the real pinning
mechanism needs further investigation.

\begin{figure}[h]
\includegraphics[width=7cm]{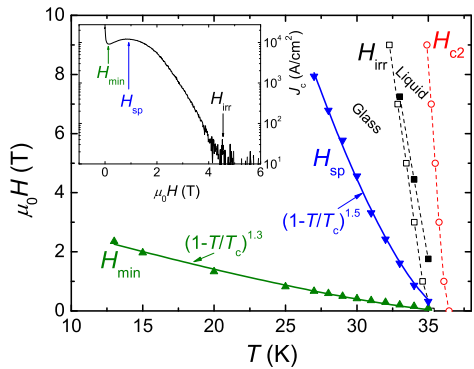}
\caption{(Color online) The phase diagram of the
Ba$_{0.6}$K$_{0.4}$Fe$_{2}$As$_{2}$ single crystal. The solid
symbols are taken from the magnetic measurements (the inset shows
the schematic diagram of the definitions of the characteristic
parameters), while the open ones are taken from the resistive
measurements. The solid lines show the fitting results to the
$H_\mathrm{min}$-$T$ and $H_\mathrm{sp}$-$T$ curves.} \label{fig5}
\end{figure}

In conclusion, we find large upper critical field and current
carrying ability as well as the fishtail effect in the single
crystal Ba$_{0.6}$K$_{0.4}$Fe$_{2}$As$_{2}$. All of them show
potential applications of the material. The fishtail effect at high
temperatures below $T_\mathrm{c}$ is originated from the small size
normal core pinning effect as the peak position of pinning force
density locates at $0.33H_\mathrm{irr}$. The vortex phase diagram of
the new material is thus obtained with magnetic field parallel to
the $c$-axis.

This work was financially supported by the Natural Science
Foundation of China, the Ministry of Science and Technology of China
(973 Projects Nos. 2006CB601000, 2006CB921802 and 2006CB921300), and
Chinese Academy of Sciences (Project ITSNEM).

$^*$Electronic mail: hhwen@aphy.iphy.ac.cn.

\end{document}